# Energy input and response from prompt and early optical afterglow emission in γ-ray bursts

W. T. Vestrand[1], J. A. Wren[1], P. R. Wozniak[1], R. Aptekar[2], S. Golentskii[2] V. Pal'shin[2], T. Sakamoto[3], R. R. White[1], S. Evans[1], D. Casperson[1] & E. Fenimore[1]

[1]Los Alamos National Laboratory, Space Science and Applications Group, ISR-1, MS-D466, Los Alamos, New Mexico 87545, USA. [2]Ioffe Physico-Technical Institute, St Petersburg, 194021, Russia. [3]NASA Goddard Space Flight Center, Code 661, Greenbelt, Maryland 20771, USA.

**The taxonomy of optical emission detected during the critical first few minutes after the onset of a γ-ray burst (GRB) defines two broad classes: prompt optical emission correlated with prompt γ-ray emission[1], and early optical afterglow emission uncorrelated with the γ-ray emission[2]. The standard theoretical interpretation attributes prompt emission to internal shocks in the ultra-relativistic outflow generated by the internal engine[3-5]; early afterglow emission is attributed to shocks generated by interaction with the surrounding medium[6-8]. Here we report on observations of a bright GRB that, for the first time, clearly show the temporal relationship and relative strength of the two optical components. The observations indicate that early afterglow emission can be understood as reverberation of the energy input measured by prompt emission. Measurements of the early afterglow reverberations therefore probe the structure of the environment around the burst, whereas the subsequent response to late-time impulsive energy releases reveals how earlier flaring episodes have altered the jet and environment parameters. Many GRBs are generated by the death of massive stars that were born and died before the Universe was ten per cent of its current age[9,10], so GRB afterglow reverberations provide clues about the environments around some of the first stars.**

On 20 August 2005 a relatively faint 30-second pulse of gamma-ray emission from GRB 050820a, which started at 06:34:53 UT, was localized[11] in real time by the autonomous software of the Burst Alert Telescope (BAT) on the Swift satellite[12]. One of our autonomous RAPTOR (**RAP**id **T**elescopes for **O**ptical **R**esponse) telescopes[16] began optical imaging of the GRB 050820a location 5.5 seconds[17] after distribution of the Swift alert. The images show emergence of faint optical emission that suddenly flares (see figure 1) and varies erratically before gradually fading over the course of the first hour. Comparison of the optical measurements with the gamma ray light curve measured by the KONUS-Wind experiment (figure 2) indicates the rapid optical flares are simultaneous with the major outbursts of gamma-ray emission. These data clearly confirm the idea that one component of the early optical light from GRBs is prompt emission that closely tracks the gamma-ray emission[1]. The remaining optical emission persists even after the gamma-rays disappear and is naturally explained as an early afterglow component.[2,19-22].

The optical light curve rise to maximum light for even the afterglow component has too much curvature (positive and negative) to be fit by the self similar power-law rise predicted by internal/external shock models. To dissect the light curve into primary components, we made the simple conjecture that the two optical components, prompt optical $F_p(t)$ and early optical afterglow $F_a(t)$ fluxes, have the temporal behaviors:

$$F_p(t) = C_p F_\gamma(t) \qquad (1)$$

$$F_a(t) = C_a \left(\frac{t-t_o}{t_o}\right)^{-s} \exp\left(\frac{-\tau}{t-t_o}\right) \qquad (2)$$

Here $F_\gamma(t)$ is the prompt gamma-ray flux; $t_o$ is the time for onset of energy release; $\tau$ is the timescale for rise of the afterglow; $s$ is the power-law decay index; and $C_p$, $C_a$ are the relative amplitudes of the prompt and early afterglow components, respectively. After re-binning the gamma-ray measurements to the same observing intervals as the optical observations (see table 1), we find that the simple temporal behaviors given by equations (1) and (2) jointly describe the optical light curve measured for GRB 050820a rather well (see figure 3).

The prompt optical component given by equation (1) reproduces the fast variations observed in the RAPTOR light curve when the KONUS gamma-ray measurements are scaled by the flux ratio $C_p = F_{opt}/F_\gamma \sim 7\times10^{-6}$---a ratio is comparable to the value, $F_{opt}/F_\gamma = 1.2\times10^{-5}$ found for GRB 041219a[1]. For both events there are intervals where the broadband prompt spectra (figure 4) between optical and gamma ray bands must on average be flatter than the $F_\nu \alpha\ \nu^{-1/2}$ expected for the standard synchrotron picture with fast cooling electrons and a cooling critical frequency near the optical band[7]. A possible explanation for closely linked, but separate, spectral components is that they are both associated with the internal jet shock, but optical emission is generated by the reverse shock and gamma rays by the forward shock[7]. This scenario would place bounds on the jet bulk velocity, the overtaking shell thickness, and the distance at which the emission originates. An alternate internal shock explanation is the optical light is synchrotron emission and the gamma rays are inverse Compton scattered synchrotron photons[4]. The closer tracking of the optical flux with the highest energy gamma-ray band (figure 4), that was also seen in GRB 041219a, is naturally explained in this scenario. Detailed modeling work will be needed to test the standard synchrotron picture and, if necessary, discriminate between possible alternative models, but the close temporal tracking of the prompt components will place important constraints on the jet properties that are difficult to get any other way.

The dramatic optical flux increase immediately after the dominant gamma-ray pulse suggests that the afterglow is associated with the impulsive energy release signaled by the prompt emission. After the rapid initial rise, the rate of flux increase slows and transitions into a shallow decline that gradually steepens to a power-law flux decay with an index of s=1.1 after ~$10^3$ seconds. Similar late time behavior has been observed for several other events [22, 23] and is consistent with that expected for an external forward shock in the internal/external shock paradigm.

While neither instrument was observing the GRB location during the precursor pulse that triggered Swift/BAT, both RAPTOR and the UVOT (Ultra-Violet and Optical Telescope) on Swift did detect[24] faint optical emission during the interval after the precursor and before the outburst of intense gamma ray emission. During that same interval neither Swift nor KONUS detected gamma ray emission. The best explanation for this faint optical emission is that it is afterglow emission from the precursor gamma-ray pulse---an explanation supported by the Swift detection of a fading soft x-ray afterglow during the same interval[25]. Further, the RAPTOR measurements of this precursor afterglow show a light curve shape that is consistent with a straightforward scaling of the main afterglow (figure 3).

Until now, observational determination of the light curve shape has been hampered by not knowing where to place the onset reference time $t_o$ [26,27]. In other words---When does the afterglow begin? Often it is placed at the burst trigger time, but it can be placed logically anywhere within, or even before, the gamma-ray emitting interval. The exact choice can substantially modify the derived shape of the early light curve[26,27]. Our observations show: (1) That the onset of the dominant afterglow component should be referenced to the time of the onset of the dominant gamma-ray pulse and (2) That the assumption of a single onset time for the afterglow, $t_o$, is too simple. The structure of the afterglow is better understood as a reverberation that is relatable to the stimulus measured by the prompt emission convolved with a transfer function representing the response of the system.

There is growing evidence that the GRB engine can impulsively release energy well after the initial explosion[28, 29]. Measurements of the broad-band spectra of the prompt emission place important constraints on the evolution of the jet itself. Structure associated with the afterglow from those secondary energy releases can also emerge as the primary afterglow component fades. The timing and strength of those secondary reverberations will probe the evolution of the interaction and how the GRB environment is modified. Since long duration GRBs are known to signal the deaths of massive stars and occur at very high redshift[9,10], measurement of GRB reverberation and its temporal variations can therefore be used to map out the nurseries of the earliest stars.

Acknowledgements: The RAPTOR project is supported by the Laboratory Directed Research and Development program at Los Alamos National Laboratory. The Konus-Wind experiment is supported by the Russian Space Agency.

Correspondence and requests for materials should be addressed to W.T.V. (e-mail: vestrand@lanl.gov).

**Table 1** Simultaneous RAPTOR/KONUS measurements of early emission from GRB 0500820a.

| Interval | $t_{start}$ (s) | $t_{end}$ (s) | $\Delta t_{exp}$ (s) | $R$ (mag) | $F$ (18–1,000 keV) ($10^{-7}$ erg cm$^{-2}$) |
|---|---|---|---|---|---|
| 0  | -4.30  | 19.30  | 23.6 | —            | 25.9±1.3  |
| 1  | 33.72  | 61.87  | 30   | >18.38       | <3.3      |
| 2  | 70.08  | 98.27  | 30   | 18.270±0.290 | <4.5      |
| 3  | 105.97 | 133.56 | 30   | >18.39       | <6.0      |
| 4  | 141.16 | 168.76 | 20   | 17.540±0.150 | <5.6      |
| 5  | 176.35 | 186.35 | 10   | 17.530±0.200 | <2.2      |
| 6  | 194.05 | 204.05 | 10   | >17.67       | <3.7      |
| 7  | 214.85 | 244.85 | 30   | 15.437±0.025 | 101.9±1.4 |
| 8  | 252.44 | 282.44 | 30   | 15.297±0.024 | 160.2±1.5 |
| 9  | 290.04 | 320.04 | 30   | 15.870±0.033 | 16.6±1.2  |
| 10 | 327.63 | 357.63 | 30   | 15.115±0.022 | <5.1      |
| 11 | 365.23 | 395.23 | 30   | 14.972±0.020 | 7.8±1.1   |
| 12 | 420.83 | 432.83 | 30   | 14.633±0.019 | 87.7±1.2  |
| 13 | 440.42 | 470.42 | 30   | 14.779±0.019 | 27.7±1.2  |
| 14 | 478.11 | 508.11 | 30   | 14.740±0.020 | 21.5±1.1  |
| 15 | 515.71 | 545.71 | 30   | 14.718±0.019 | 51.5±1.1  |
| 16 | 553.30 | 583.30 | 30   | 14.764±0.019 | 11.5±1.1  |
| 17 | 594.20 | 624.20 | 30   | 14.901±0.020 | 4.7±3.3   |
| 18 | 631.79 | 661.79 | 30   | 14.983±0.020 | <5.4      |
| 19 | 669.39 | 699.39 | 30   | 15.038±0.021 | 11.9±1.1  |
| 20 | 706.98 | 736.98 | 30   | 15.060±0.022 | 11.5±1.3  |

The R-band magnitudes were derived by transforming the unfiltered measurements to an R-band equivalent using the USNO-B1.0 magnitudes for comparison stars in the RAPTOR images.

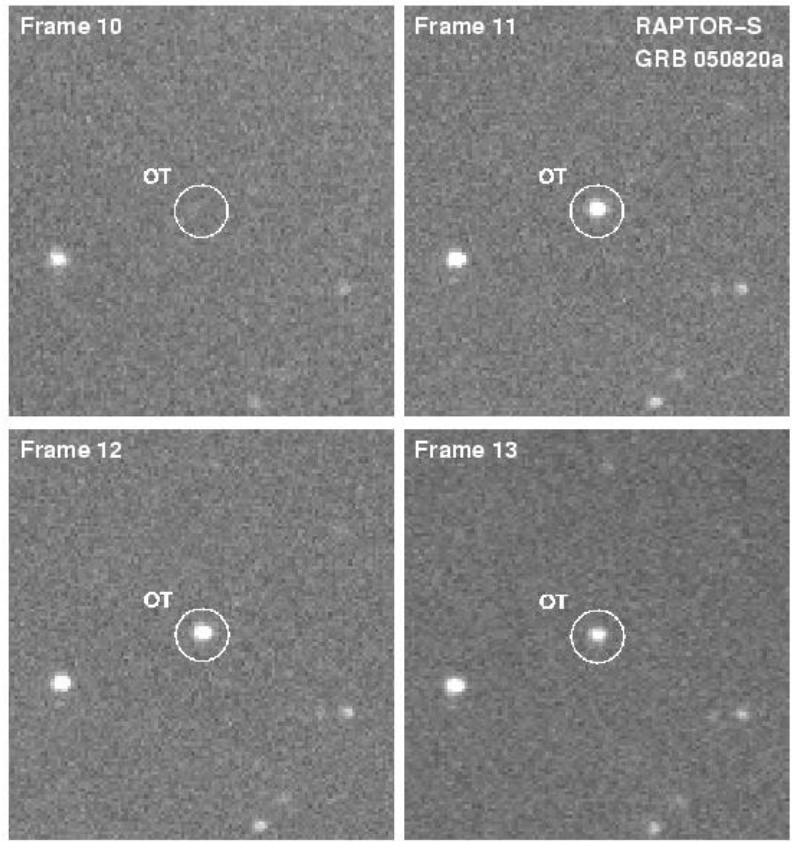

Figure 1. The onset of prompt optical emission from GRB 050820a. The images show the GRB location in four consecutive frames collected by the 0.4-meter fully autonomous rapid response telescope, RAPTOR-S, owned by Los Alamos National Laboratory and located at an altitude of 2,500 m in the Jemez Mountains of New Mexico. The RAPTOR-S telescope employs an unfiltered 1Kx1K pixel back-illuminated CCD camera and typically achieves a 5-sigma limiting magnitude of R~19th magnitude for 30-second exposures. The displayed images span the time interval 06:38:07.2 to 06:40:50.7 UT on 2005 August 20 and the circle denotes the location of the GRB optical counterpart (right ascension 22h 29 min 38.1s, declination +19°33'37.1" (J2000))[18] that subsequent spectral measurements[14,15] showed was at a redshift of z=2.612+/-0.002. The top two images (Frames 10 and 11) show flaring of the optical transient (OT) by more than two magnitudes in less than 30 seconds. The bottom two images (Frames 12 and 13) are the next two 30-second exposures which show an abrupt drop of the optical emission in Frame 13 by ~ 0.5 magnitudes.

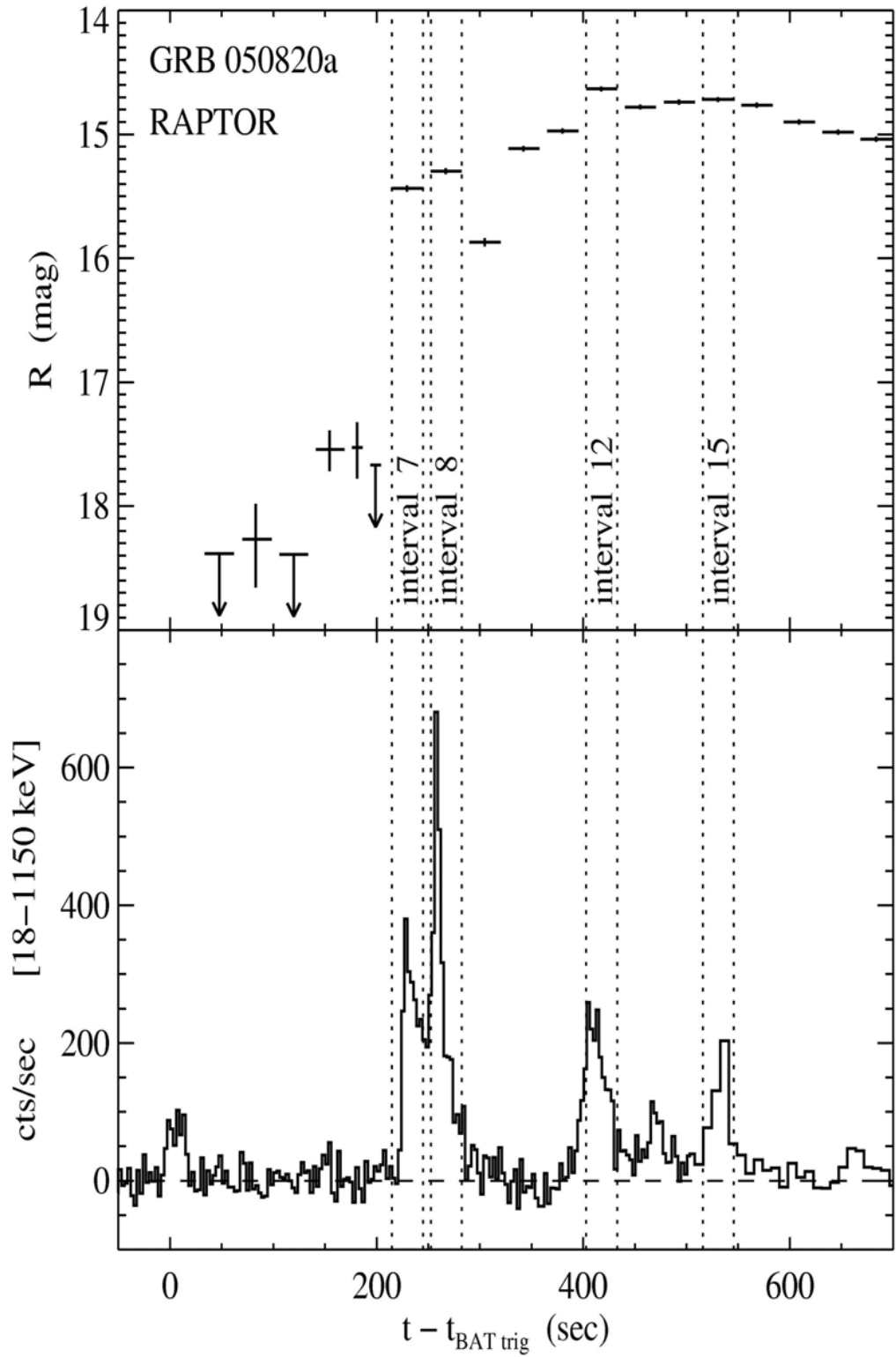

Figure 2. A comparison of the early optical light curve and the gamma-ray light curve measured for GRB 050820a. The lower panel shows the gamma-ray count rate in the light curve measured by KONUS gamma-ray detectors on board the WIND satellite. The upper panel shows the optical light curve measured by the RAPTOR-S telescope.

After that initial pulse, the GRB was quiescent at gamma-ray energies for more than three minutes until a major outburst began at 06:38:40 UT. The observations by Swift were truncated by satellite passage into a high background region, but the KONUS gamma-ray detector aboard the Wind satellite was able to measure the entire complex gamma-ray light curve[13] that lasted ~750 seconds. Integrated over the entire event, the total (18-1000 keV) gamma-ray fluence for GRB 050820a was $5.3 \times 10^{-5}$ erg/cm$^2$, which for the observed redshift [14,15] (z=2.6) and standard cosmological parameters ($\Omega_m$=0.3, $\Omega_\Lambda$=0.7, $H_0$=70 km/s/Mpc) corresponds to an isotropic energy of $8.1 \times 10^{53}$ ergs. The abrupt brightening of the optical emission shown in figure 1 occurs simultaneously with the major pulses of gamma-ray emission shown in the interval between 226 and 290 seconds after the burst trigger. The onset of early optical afterglow emission is visible in the interval between 300 and 400 seconds after the burst trigger.

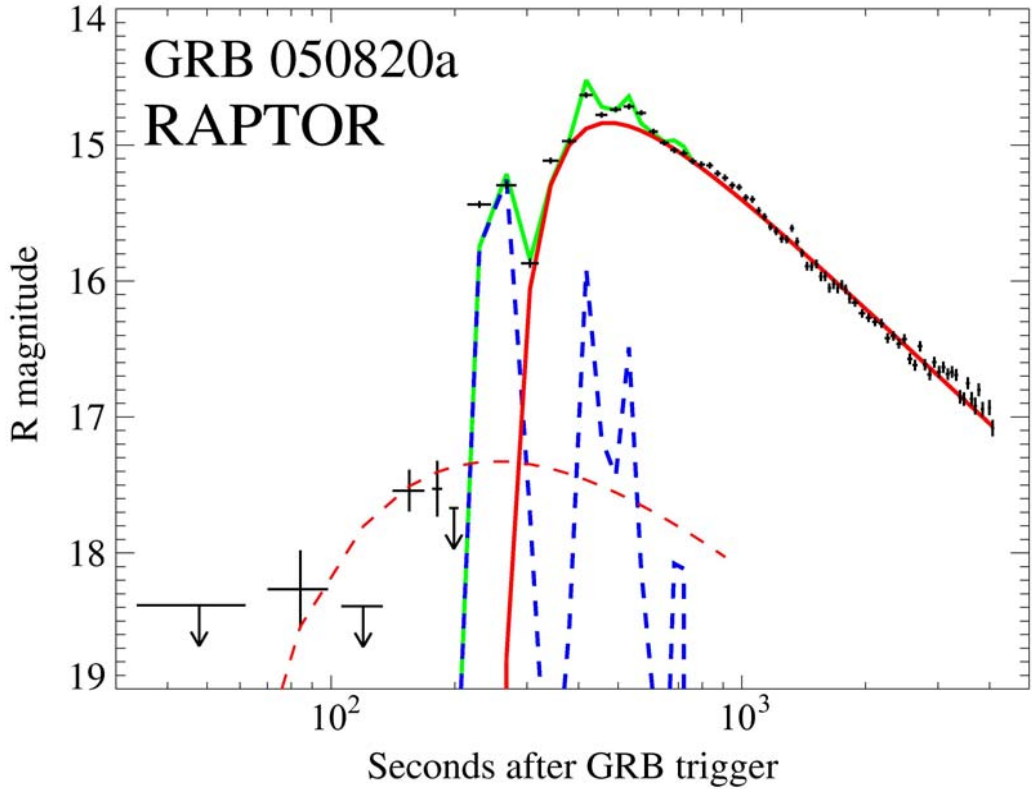

Figure 3. The decomposition the optical light curve measured for GRB 050820a into primary optical components. The R-band magnitudes measured by RAPTOR-S are indicated by the black crosses, with observing intervals denoted by horizontal lines and 1-sigma error bars represented by the vertical lines. The blue dashed trace shows the prompt optical component obtained by scaling the KONUS gamma-ray flux measurements, after re-binning to the same time intervals as the RAPTOR measurements, by the factor $F_{opt}/F_\gamma = 7.4 \times 10^{-6}$ and converting to the R-band equivalent magnitude. The solid red line shows the model early afterglow component of the form given by equation (2) with a flux rise timescale of $\tau \approx 280$ seconds, late time power-law flux decay with index s~1.1, and a reference time $t_o$ equal to the start of the dominant gamma-ray outburst (226 seconds after the BAT trigger time) . The green trace shows the sum of these prompt and early afterglow model components. The dashed red line shows an afterglow with the same flux rise timescale as the dominant afterglow

component ($\tau \approx$ 280 seconds) but with a reference time appropriate for the precursor pulse ($t_o$=5 seconds) and an amplitude given by the ratio of the gamma-ray fluences for the precursor pulse and the main pulse (precursor/main pulse (18-1150 keV) fluence=0.1). Note that even when the prompt gamma ray emitting intervals are dropped, the rise of the afterglow emission cannot be fit with a power-law dependence. The fitting of a power law with a reference time of $t_o$=0.0 to the fast rise in the interval measured with unprecedented signal-to-noise between 300 and 400 seconds after the trigger, predicts fluxes that are inconsistent with the detections of optical emission by RAPTOR and the UVOT in the interval before 200 seconds after the trigger.

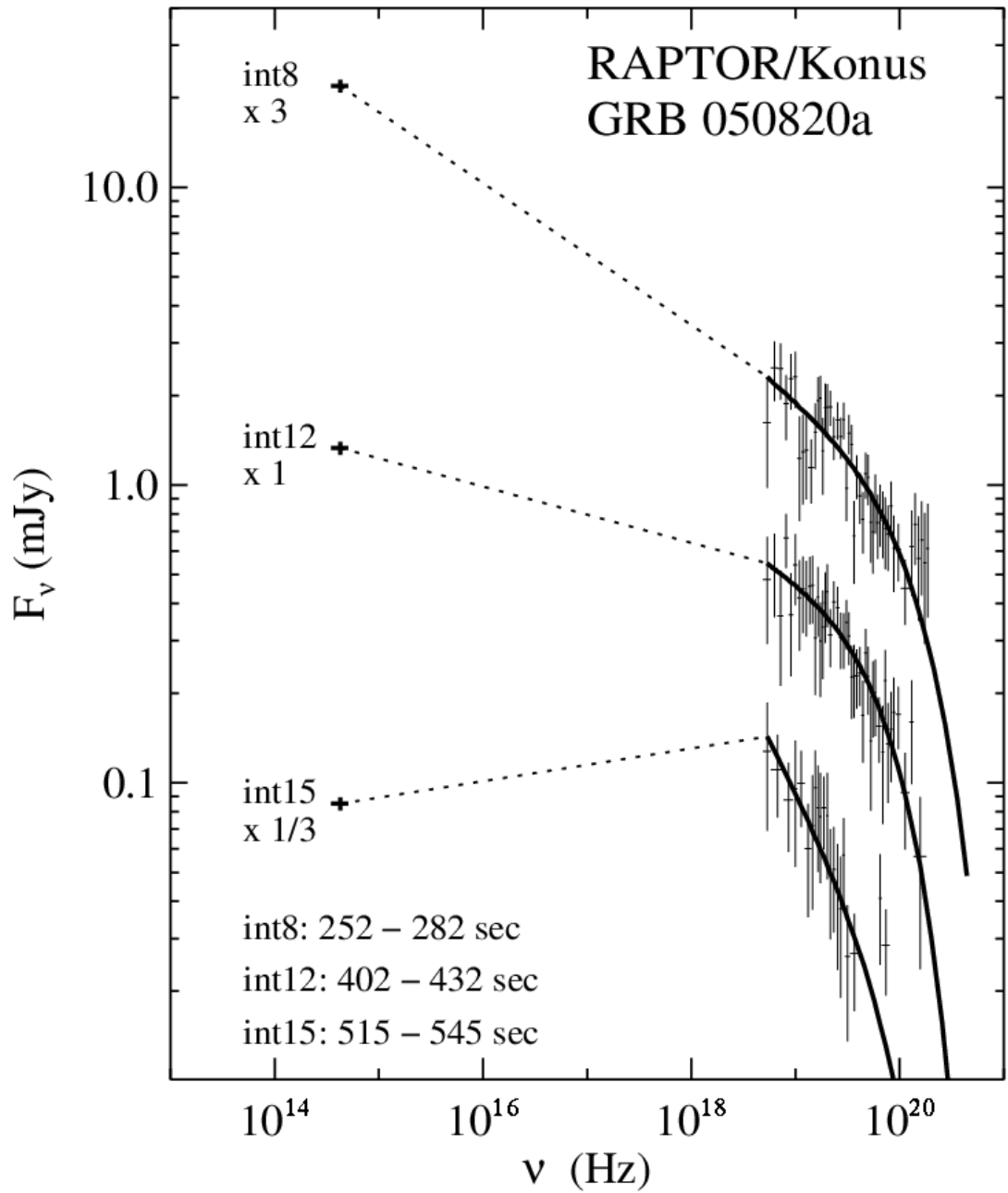

Figure 4. Broad-band spectra of prompt emission from GRB 050820a. These spectra are constructed from simultaneous optical measurements from the RAPTOR-S telescope and gamma-ray measurements from the KONUS experiment on board the WIND Satellite. The crosses at the low frequencies denote the optical flux density after correction for the contribution from early optical afterglow emission. The solid line denotes the best fitting model for the KONUS measurements and the high frequency crosses represent the individual channel measurements with the 1-sigma errors. The time intervals are measured in seconds from the Swift trigger time. During intervals 8 and 12, the broadband spectra of prompt emission for GRB 050820a show optical flux density levels that are roughly consistent with an extrapolation of the high energy spectral shape. But in interval 15, extrapolation of the gamma ray spectrum significantly under predicts the optical flux. Notice that the highest energy gamma ray flux seems to be a slightly better predictor of the behavior of the prompt optical flux---a behavior also observed[1] in GRB 041219a. This suggests that the prompt optical and gamma-ray emission might be generated by separate, but closely linked, radiation processes.